\documentclass[10pt,oneside,twocolumn,a4paper]{article}
\usepackage{graphicx}
\usepackage[utf8]{inputenc}
\usepackage{mathptmx}
\usepackage{tabularx}
\usepackage{ragged2e}
\usepackage[singlelinecheck=false]{caption}
\usepackage[T1]{fontenc}
\usepackage[numbers]{natbib}
\usepackage{amsmath}
\usepackage{textcomp}
\usepackage{calc}
\usepackage{ifthen}
\usepackage{tikz}
\usetikzlibrary{automata, positioning, arrows}

\tikzset{
    ->, % makes the edges directed
    >=stealth,
    node distance=4em, % specifies the minimum distance between two nodes. Change if necessary.
    every state/.style={thick, fill=gray!10, minimum size=4em}, % sets the properties for each ’state’ node
    initial text=$ $, % sets the text that appears on the start arrow
    fontscale/.style = {font=\scriptsize}
}

\definecolor{ownblue}{rgb}{0.0, 0.59, 0.85}

\RequirePackage[blocks]{authblk}
\usepackage{babel}
\makeatletter
\let\ps@plain\ps@empty
\def\@xivpt{14bp}

\def\@sect#1#2#3#4#5#6[#7]#8{%
  \ifnum #2>\c@secnumdepth
    \let\@svsec\@empty
  \else
    \refstepcounter{#1}%
    \protected@edef\@svsec{%
      \ifnum #2<4
        \hb@xt@10mm{\csname the#1\endcsname}\relax
      \else
        \hb@xt@12mm{\csname the#1\endcsname}\relax
      \fi}%
  \fi
  \@tempskipa #5\relax
  \ifdim \@tempskipa>\z@
    \begingroup
      #6{%
        \@hangfrom{\hskip #3\relax\@svsec}%
          \interlinepenalty \@M #8\@@par}%
    \endgroup
    \csname #1mark\endcsname{#7}%
    \addcontentsline{toc}{#1}{%
      \ifnum #2>\c@secnumdepth \else
        \protect\numberline{\csname the#1\endcsname}%
      \fi
      #7}%
  \else
    \def\@svsechd{%
      #6{\hskip #3\relax
      \@svsec #8}%
      \csname #1mark\endcsname{#7}%
      \addcontentsline{toc}{#1}{%
        \ifnum #2>\c@secnumdepth \else
          \protect\numberline{\csname the#1\endcsname}%
        \fi
        #7}}%
  \fi
  \@xsect{#5}}
\renewcommand\LARGE{\@setfontsize\LARGE{16}{20}}
\def\abstract#1{\def\@abstract{#1}}
\def\abstractEn#1{\def\@abstractEn{#1}}
\def\titleEn#1{\def\@titleEn{#1}}
\def\@maketitle{%
  \newpage
  \null
  \let \footnote \thanks
    {\LARGE\bfseries\RaggedRight \@title \par}%
    {\LARGE\bfseries\RaggedRight \@titleEn \par}%
    \vskip 1\baselineskip%
    {\normalsize
      \@author\par}%
    \vskip \baselineskip%
    {\section*{Abstract}
      \@abstractEn}%
  \par
  \vskip 3\baselineskip}

\renewcommand\section{\@startsection {section}{1}{\z@}%
                                   {-3.5ex \@plus -1ex \@minus -.2ex}%
                                   {\baselineskip}%
                                   {\normalfont\Large\bfseries\RaggedRight}}
\renewcommand\subsection{\@startsection{subsection}{2}{\z@}%
                                     {\baselineskip}%
                                     {1ex}%
                                     {\normalfont\large\bfseries\RaggedRight}}
\renewcommand\subsubsection{\@startsection{subsubsection}{3}{\z@}%
                                     {1\baselineskip}%
                                     {3bp}%
                                     {\normalfont\normalsize\bfseries\RaggedRight}}
\renewcommand\paragraph{\@startsection{paragraph}{4}{\z@}%
                                    {1\baselineskip\@plus1ex \@minus.2ex}%
                                    {3bp}%
                                    {\normalfont\normalsize\RaggedRight}}
\renewcommand\subparagraph{\@startsection{subparagraph}{5}{\parindent}%
                                       {3.25ex \@plus1ex \@minus .2ex}%
                                       {-1em}%
                                      {\normalfont\normalsize\bfseries\RaggedRight}}
\parindent\p@
\makeatother
\DeclareCaptionLabelSeparator{enskip}{\enskip}
\DeclareFixedFont{\ttb}{T1}{txtt}{bx}{n}{12} % for bold
\DeclareFixedFont{\ttm}{T1}{txtt}{m}{n}{12}  % for normal

\usepackage{color}
\definecolor{deepblue}{rgb}{0,0,0.5}
\definecolor{deepred}{rgb}{0.6,0,0}
\definecolor{deepgreen}{rgb}{0,0.5,0}

\usepackage{listings}

\usepackage{xcolor}

\definecolor{codegreen}{rgb}{0.24, 0.47, 0.35}
\definecolor{codegray}{rgb}{0.5,0.5,0.5}
\definecolor{codepurple}{rgb}{0.58,0,0.82}
\definecolor{codeblue}{rgb}{0.58,0,0.82}
\definecolor{backcolour}{rgb}{0.95,0.95,0.92}

\lstdefinestyle{mystyle}{
    commentstyle=\color{codegreen},
    keywordstyle=\color{blue},
    numberstyle=\tiny,
    stringstyle=\color{codepurple},
    basicstyle=\ttfamily\footnotesize,
    breakatwhitespace=false,         
    breaklines=true,                 
    captionpos=b,                    
    keepspaces=true,                 
    numbers=left,                    
    numbersep=5pt,                  
    showspaces=false,                
    showstringspaces=false,
    showtabs=false,                  
    tabsize=2,
    floatplacement=tbp
}

\newcommand{\goenna}{$\ddot{\Gamma}$}

\newcounter{captionedequationset} %numbering
\newdimen\captionlength
\newcommand{\captionedequationset}[1]{
    \refstepcounter{captionedequationset}% Step counter
    \setlength{\captionlength}{\widthof{#1}} %
    \addtolength{\captionlength}{\widthof{Equation set~\thecaptionedequationset: }}
    \ifthenelse{\lengthtest{\captionlength < \linewidth }} %
    {\begin{center}
            Equation set~\thecaptionedequationset: #1
        \end{center}} 
    { \begin{flushleft} 
        Equation set~\thecaptionedequationset: #1 %
        \end{flushleft}}}

\title{Using the Abstract Computer Architecture Description Language to Model AI Hardware Accelerators}
\titleEn{}
\author{Mika Markus Müller\textsuperscript{*}, Alexander Richard Manfred Borst\textsuperscript{*}, Konstantin Lübeck\textsuperscript{*}, Alexander Louis-Ferdinand Jung, and Oliver Bringmann}
\affil{Embedded Systems, Department of Computer Science\\Eberhard Karls Universität Tübingen, Tübingen, Germany}
\affil{\texttt{\{mika.mueller, a.borst, konstantin.luebeck, a.jung, oliver.bringmann\}@uni-tuebingen.de}}
\affil{\newline\textsuperscript{*}These authors contributed equally to this work.}

\abstractEn{Artificial Intelligence (AI) has witnessed remarkable growth, particularly through the proliferation of Deep Neural Networks (DNNs). These powerful models drive technological advancements across various domains. However, to harness their potential in real-world applications, specialized hardware accelerators are essential. This demand has sparked a market for parameterizable AI hardware accelerators offered by different vendors.\\
Manufacturers of AI-integrated products face a critical challenge: selecting an accelerator that aligns with their product’s performance requirements. The decision involves choosing the right hardware and configuring a suitable set of parameters. However, comparing different accelerator design alternatives remains a complex task. Often, engineers rely on data sheets, spreadsheet calculations, or slow black-box simulators, which only offer a coarse understanding of the performance characteristics.\\
The Abstract Computer Architecture Description Language (ACADL) is a concise formalization of computer architecture block diagrams, which helps to communicate computer architecture on different abstraction levels and allows for inferring performance characteristics. In this paper, we demonstrate how to use the ACADL to model AI hardware accelerators, use their ACADL description to map DNNs onto them, and explain the timing simulation semantics to gather performance results.}

\begin{document}

\maketitle

\section{Introduction}
The recent advances in Artificial Intelligence (AI) have resulted in an exponential increase in performance and energy demands for the inference of Deep Neural Networks (DNN). To keep up with those increased demands, specialized AI hardware accelerators are needed, which offer better performance and energy-efficiency compared to CPUs and general-purpose GPUs \cite{dccnn2010,fpgavsgpu2017,crc2019,microsoft2023}. However, evaluating and comparing different AI hardware accelerator architectures is a time-consuming task that often relies on simulators with poor simulation performance or data sheets with questionable performance claims. The Abstract Computer Architecture Description Language (ACADL) provides a concise modeling of arbitrary computer architectures along with a timing semantic that allows for inferring performance characteristics from modeled AI hardware accelerators.

In this paper, we present how to use ACADL to model AI hardware accelerators on different abstraction levels and how to map DNN operators onto those models to infer performance characteristics to speed-up accelerator selection and design, hardware-aware Neural Architecture Search (NAS), and DNN/HW Co-Design \cite{bestofbothworlds2020, jiang2020, zhou2021, gerum2022}.

This paper makes the following contributions:

\begin{itemize}
    \item An introduction to ACADL,
    \newpage
    \item practical examples that demonstrate how to use ACADL to model AI hardware accelerators,
    \item an overview on how to map DNNs on to the AI hardware accelerators described in ACADL,
    \item and, an explanation of the timing simulation semantics to infer performance metrics from DNN mappings.
\end{itemize}

\section{Related Work}
Several methods for modeling AI hardware accelerators and general computer architectures for performance estimation have been proposed. 

ScaleSim \cite{scalesim2019} presents a DNN accelerator simulation framework based on a coarse accelerator template. This accelerator template is parameterizable using a set of ten parameters which specify the number of multiply accumulate units, the memory sizes, the data flow, etc. The authors conduct several experiments showing the performance and energy estimations for a range of different DNNs and hardware configurations.

Timeloop \cite{timeloop2019} introduces a domain-specific language for DNN accelerators, focusing on the memory hierarchy and the supported data flows. They validate their proposed approach by modeling three real-world accelerator architectures and estimating their performance and power.

DNN-Chip Predictor \cite{dnnchippredictor2020} also proposes a domain-specific language which expresses how a DNN layer is computed by a hardware accelerator using nested loops and memory hierarchy. The authors validate their modeling and estimation framework with an FPGA implementation.

SimPyler \cite{simpyler2023} provides a Python-based methodology for composing hardware templates with embedded analytical models into complete accelerator platforms. Using the TVM \cite{tvm2018} compiler framework, the authors can transform and map DNN operators onto an accelerator platform, which is then simulated as a Petri-Net calling the analytical models to infer performance results.

HDPython \cite{hdpython2023} is a Python-based Hardware Description Language (HDL) which uses Python as a source language which is then converted into VHDL. While HDPython facilitates object-orientation and generic programming, it does not allow the analysis of the described hardware on an abstraction level above the register-transfer level.

IP-XACT \cite{ipxact2008} provides a standard for the integration and verification of intellectual property (IP) in the context of System-on-Chip (SoC) designs. IP-XACT enables the use of virtual prototypes for early design phases, including software compilation and debugging, which allows for the integration of AI hardware accelerator models.

While the works mentioned above either use register-transfer implementations, parameterizable accelerator templates, or introduce a domain-specific language, this paper presents a flexible modeling methodology for AI hardware accelerators implemented in C++ with a Python front-end that enables modeling at different operator abstraction levels for arbitrary computer architectures without relying on proprietary software and complex tool flows.

\section{Abstract Computer Architecture Description Language}
\label{sec:acadl}
\begin{figure*}[th]
    \centering
    \includegraphics[width=0.98\textwidth]{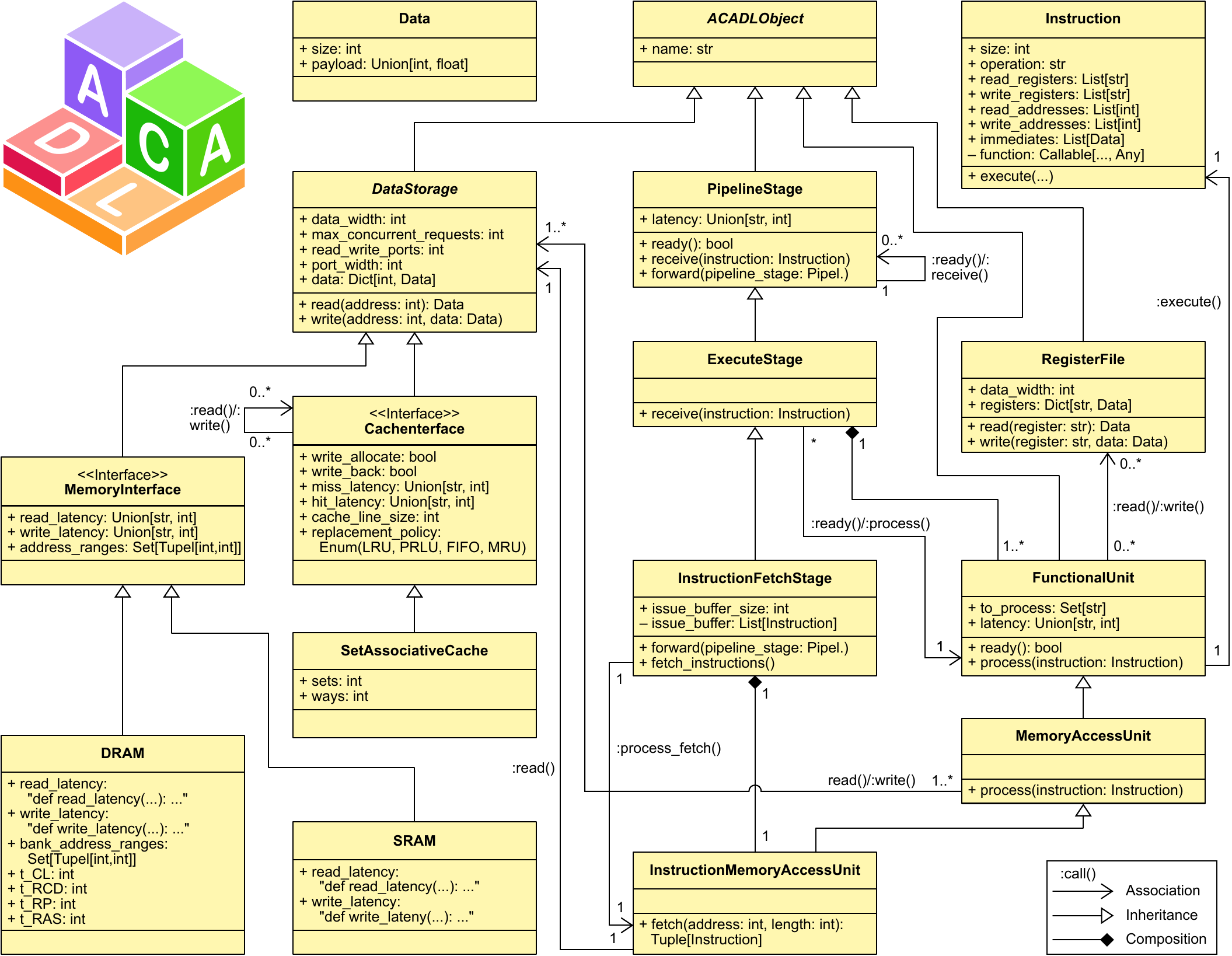}
    \caption{Class diagram of the Abstract Computer Architecture Description Language.}
    \label{fig:acadl_class_diagram}
\end{figure*}

Computer Architectures are almost exclusively communicated using block diagrams, where blocks describe how data is transformed and edges signify dependencies and data flow between different blocks. However, those computer architecture block diagrams are often ambiguous and therefore need textual explanation to be understood, which forbids an automated analysis. The Abstract Computer Architecture Description Language (ACADL) \cite{acadl2022} solves this problem by formalizing computer architecture block diagrams by providing a set of basic building blocks and their relations together with a concise modeling and timing semantic, enabling both human and machine-readable representations, allowing higher-level system descriptions.

Using Prabhat's and Dutt's taxonomy \cite{pdls2008} ACADL can be classified as a behavioral architecture description language (ADL). ACADL ignores register-transfer and gate-level details and focuses on the timing semantics of operations on the scalar, tensor, and fused-tensor abstraction level using an object-oriented paradigm.

ACADL is an instruction-centric ADL, which means that any architectural state change is triggered by an instruction being processed. ACADL consists of a virtual base class, twelve classes, and two interfaces as presented in the UML \cite{uml2017} class diagram in Fig.~\ref{fig:acadl_class_diagram}. ACADL classes are instantiated into objects when building computer architecture models, which is presented in section \ref{sec:moodeling_examples}.

The following section provides a brief overview of all ACADL elements:

\textbf{bool, int, str, List, Set, Union, Tuple, Any, Callable, Enum, ...} are the primitive types of ACADL based on the \mbox{Python 3.10} typing system \cite{python3typing2023}.

\textbf{latency} describes a time delta in clock cycles. It can be specified as an integer value or a string containing a function that is evaluated during the performance estimation.

\textbf{Association} describes the relation between two or more instantiated objects of classes that allows one object (caller) to cause another one to perform an action (callee), while \textbf{:call()} signifies the callee's function that is called by the caller.

\textbf{Inheritance} allows a class to base its implementation and attributes upon another class (base class).

\textbf{Composition} defines a relationship between instantiated objects of classes that implies that one or more objects are part of a single composite object.

\textbf{ACADLObject} is the virtual base class for every computer architecture module modeled in ACADL. It only has the attribute \texttt{name}, the unique identifier for each object.

\textbf{Data} represents any data stored in memories, registers, and immediate values of instructions. \texttt{size} is the data size in bits. \texttt{payload} is the data itself, which is used for the functional simulation.

\textbf{Instruction} has several attributes describing which registers (\texttt{read\_registers}, \texttt{write\_registers}) and memory addresses (\texttt{read\_addresses}, \texttt{write\_addresses}) and immediate values (\texttt{immediates}) are accessed when it is executed. \texttt{operation} is the instruction's mnemonic, while \texttt{function} contains how data is manipulated. \texttt{execute()} calls \texttt{function} when the instruction is processed by a FunctionalUnit. Furthermore, an instruction is not limited to fine-grained operations such as addition or multiplication. An instruction can also carry out complex operations like matrix-matrix multiplication or Fourier transformations. This enables modeling at different abstraction levels.

\textbf{PipelineStage} is responsible for forwarding instructions inside a computer architecture. \texttt{receive()} is called by \texttt{forward()} of another PipelineStage which forwards an instruction between two PipelineStages. An Instruction can only be forwarded if the receiving PipelineStage is \texttt{ready()}. The amount of clock cycles an Instruction resides inside a PipelineStage before it is forwarded is indicated by \texttt{latency}.

\textbf{RegisterFile} contains \texttt{registers} that maps the unique register name to a value. \texttt{data\_width} is the size of each register in bits, while \texttt{read()} and \texttt{write()} provide access to the stored values.

\textbf{FunctionalUnit} executes an Instruction that is passed to \texttt{process()} and changes the architectural state by changing register contents using \texttt{read()} and \texttt{write()} of RegisterFiles. A RegisterFile can be read-only, write-only, or readable and writable depending on the \texttt{:read()}/\texttt{:write()} associations connecting FunctionalUnits and RegisterFiles. FunctionalUnits can only process Instructions whose \texttt{operation} is in \texttt{to\_process} and if the FunctionalUnit has read/write access to the \texttt{read\_registers} and \texttt{write\_registers}. Processing a supported Instruction takes \texttt{latency} clock cycles after all data dependencies from previous Instructions are resolved.

\textbf{ExecuteStage} inherits from PipelineStage and overrides \texttt{receive()} and additionally contains FunctionalUnits. When an ExecuteStage receives an Instruction, it checks all its contained FunctionalUnits if one of them supports processing the Instruction. The check involves if \texttt{operation} is in \texttt{to\_process} and if the \texttt{read\_registers} and \texttt{write\_registers} are accessible by the FunctionalUnit. When a contained FunctionalUnit supports the Instruction, the ExecuteStage passes the Instruction to \texttt{process()} and the ExecuteStage's \texttt{latency} is not accumulated. 

\textbf{DataStorage} is the virtual base class for all data storages. \texttt{data\_width} describes the bit-length of one data word. \texttt{max\_concurrent\_requests} is the maximum amount of read and write requests that can be handled at the same time, while \texttt{read\_write\_ports} describes how many MemoryAccessUnits can be connected to the DataStorage. \texttt{port\_width} is the amount of data words that can be accessed during a single memory transaction. A $\texttt{port\_width} > 1$ allows for reading or writing several data words at once. \texttt{data} maps a memory address to a data word.

\textbf{MemoryInterface} adds \texttt{read\_latency}, \texttt{write\_latency}, and \texttt{address\_ranges} to DataStorage which describe how many clock cycles a read or write transaction takes when \texttt{read()} or \texttt{write()} are called and which addresses are associated with a memory. 

\textbf{DRAM} and \textbf{SRAM} override \texttt{read\_latency} and \texttt{write\_latency} with stateful functions for accurate memory timing simulation. Additionally, DRAM contains the attributes \texttt{bank\_address\_ranges}, \texttt{t\_RCD}, \texttt{t\_RP}, and \texttt{t\_RAS} which are used by the latency functions.

\textbf{CacheInterface} adds \texttt{write\_allocate}, \texttt{write\_back}, \texttt{miss\_latency}, \texttt{hit\_latency}, \texttt{cache\_line\_size}, and \texttt{replacement\_policy} to DataStorage which are common attributes of a cache.

\textbf{SetAssociativeCache} extends CacheInterface and adds \texttt{sets} and \texttt{ways}.

\textbf{MemoryAccessUnit} inherits from FunctionalUnit and overrides \texttt{process()} to access RegisterFiles and objects that inherit from DataStorage. 
\newpage
\textbf{InstructionMemoryAccessUnit} inherits from MemoryAccessUnit and adds \texttt{fetch()}, which reads \texttt{length} Instructions starting at \texttt{address} from a DataStorage that contains instructions, which we call instruction memory for the remainder of this paper.

\textbf{InstructionFetchStage} inherits from ExecuteStage and overrides \texttt{forward()}, which forwards an instruction residing in the \texttt{issue\_buffer}. It is possible to forward multiple Instructions in the same clock cycle. \texttt{issue\_buffer\_size} describes how many instructions fit into the \texttt{issue\_buffer}, and thereby marks the maximum number of instructions that can be issued in a single clock cycle. \texttt{fetch\_instructions()} is called in each clock cycle as long as the \texttt{issue\_buffer} is not full.

Using those classes and their dependencies, a wide variety of computer architectures can be modeled using ACADL. ACADL's object-oriented design facilitates expendability and composition. We implemented ACADL in C++ with a Python front-end and a functional instruction set simulation, which enables a seamless interaction with popular AI frameworks and tool flows such as PyTorch \cite{pytorch2019} and TVM \cite{tvm2018}. We chose Python as a front-end language because, according to the "TIOBE Index for January 2024" \cite{tiobeindex2024} Python is the most popular programming language, and according to the "Stack Overflow 2023 Developer Survey" \cite{stackoverflowsurvey2023} is placed third as the most commonly used programming language among developers. This combination of formalization and practical usability empowers researchers and engineers to accurately describe and analyze AI hardware accelerators and make informed decisions.

\section{Modeling Examples}
\label{sec:moodeling_examples}
To model a computer architecture using ACADL, objects of the twelve classes must be instantiated and connected to each other according to the class diagram depicted in Fig.~\ref{fig:acadl_class_diagram}. For the remainder of this paper, we call the UML object diagram describing a computer architecture the architecture graph (AG). In the following sections, modeling examples for different AI hardware accelerator architectures are shown for different abstraction levels, starting at the scalar operations level and ending with fused-tensor operations.

\subsection{One MAC Accelerator}
\begin{figure}[tb]
    \centering
    \includegraphics[width=0.90\linewidth]{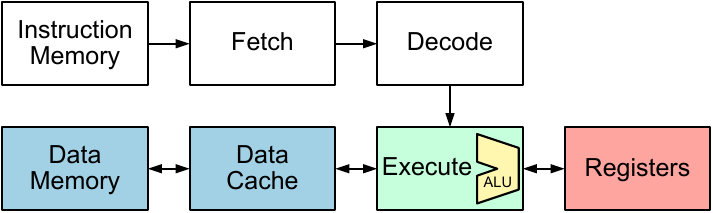}
    \caption{Block diagram of the One MAC Accelerator (OMA).}
    \label{fig:one_mac_accelerator_block_diagram}
\end{figure}

As an introductory example, we present the One MAC Accelerator (OMA) which is modeled on the scalar operations level. It is designed to accelerate the execution of DNN models using the built-in multiply-accumulate (MAC) operation. The OMA is composed out of a single data memory, a data cache, a register file, and an execution unit responsible for data manipulation and memory access. Furthermore, the OMA is only capable of processing one operation at a time in its execution unit. Fig.~\ref{fig:one_mac_accelerator_block_diagram} shows the block diagram of the One MAC Accelerator.

\begin{figure}[tb]
    \centering
    \includegraphics[width=\linewidth]{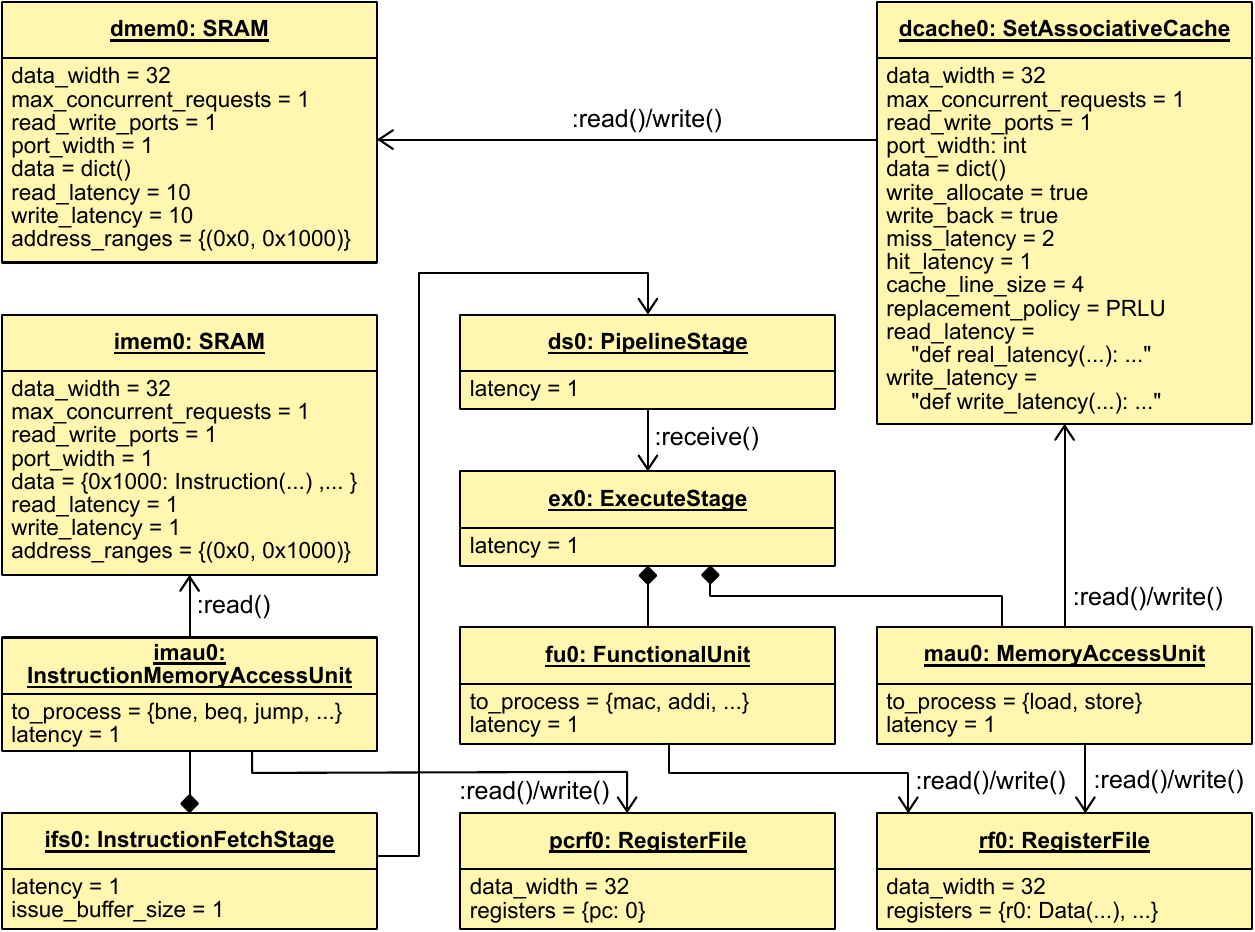}
    \caption{Architecture graph of the One MAC Accelerator (OMA).}
    \label{fig:one_mac_accelerator_object_diagram}
\end{figure}

From the OMA block diagram, we can construct the AG in Fig.~\ref{fig:one_mac_accelerator_object_diagram}. All hardware elements from the block diagram can be found in the AG. Additionally, the AG contains a RegisterFile for the program counter register \texttt{pc} and the fetch stage is split into an InstructionFetchStage \texttt{imau0} and an InstructionMemoryAccessUnit \texttt{ifs0} while the arithmetic logic unit is represented by the FunctionalUnit \texttt{fu0} and the MemoryAccessUnit \texttt{mau0}.

We can use the ACADL Python front-end to implement the AG of a computer architecture. 

Listing~\ref{lst:one_mac_accelerator_implementation} shows the Python implementation of the OMA. Each computer architecture implementation is encapsulated in a Python function decorated with \texttt{@generate}, which implicitly checks the validity of all edges when the function is called. Afterward, the ACADL built-in function \texttt{create\_ag()} is called, which instantiates the AG of the modeled architecture. The instantiation of the hardware elements as ACADL objects is straightforward. To connect the instantiated objects, the class \texttt{ACADLEdge} is used, which takes the references of objects and the edge type.

To execute operations on the OMA, an operator implementation in the form of ACADL instructions is needed. Listing~\ref{lst:gemm_example} presents an example of ACADL instructions implementing a general matrix multiplication on the OMA architecture, which is detailed in section~\ref{sec:operator_mapping}. 

\begin{lstlisting}[language=Python, caption=ACADL Python front-end implementation of the One MAC Accelerator (OMA)., label=lst:one_mac_accelerator_implementation]
from acadl import *

@generate
def generate_architecture()
    # instruction fetch
    imem0 = SRAM(name="imem0", ...)
    imau0 = InstructionMemoryAccessUnit(
        name="imau0", ...)
    pcrf0 = RegisterFile(name="pcrf0", ...)
    ifs0 = InstructionFetchStage(
        name="ifs0", ...)
    # instruction processing
    ds0 = PipelineStage(name="ds0", ...)
    ex0 = ExecuteStage(name="ex", ...)
    fu0 = FunctionalUnit(
        name="fu0", 
        to_process={"mov", "addi", ...}, 
        latency=latency_t(1)
    )
    mau0 = MemoryAccessUnit(
        name="mau0",
        to_process={"load", "store"},
        latency=latency_t(1)
    )
    rf0 = RegisterFile(
        name="rf0",
        data_width=32,
        registers={"r0": Data(32, 0), ...}
    )
    dmem0 = SRAM(name="dmem0", ...)
    dcache0 = SetAssociativeCache(
        name="dcache0", ...
    )
    
    # edges
    ACADLEdge(imem0, imau0, READ_DATA)
    ACADLEdge(pcrf0, imau0, READ_DATA)
    ACADLEdge(imau0, pcrf0, WRITE_DATA)
    ACADLEdge(ifs0, imau0, CONTAINS)
    ACADLEdge(ifs0, ds0, FORWARD)
    ACADLEdge(ds0, ex0, FORWARD)
    ACADLEdge(ex0, fu0, CONTAINS)
    ACADLEdge(fu0, rf0, WRITE_DATA)
    ACADLEdge(rf0, fu0, READ_DATA)
    ACADLEdge(ex0, mau0, CONTAINS)
    ACADLEdge(mau0, rf0, WRITE_DATA)
    ACADLEdge(rf0, mau0, READ_DATA)
    ACADLEdge(mau0, dcache0, WRITE_DATA)
    ACADLEdge(dcache0, mau0, READ_DATA)
    ACADLEdge(dcache0, dmem0, WRITE_DATA)
    ACADLEdge(dmem0, dcache0, READ_DATA)

def __name__ == "__main__":
    generate_architecture()
    ag = create_ag()
\end{lstlisting}

\subsection{Parameterizable Systolic Array}
While modeling the OMA is straightforward, ACADL also allows modeling more complex and flexible architectures whose structure depends on specific parameters. In the following, we demonstrate this capability by modeling a parameterizable systolic array using ACADL. 

\begin{figure}[tb]
    \centering
    \includegraphics[width=0.9\linewidth]{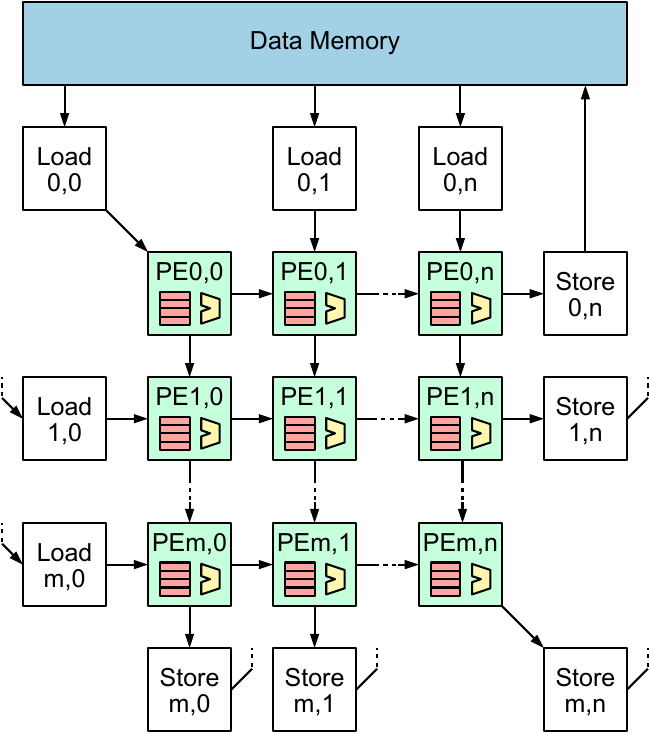}
    \caption{Block diagram of the systolic array architecture (excluding instruction memory and fetch unit).}
    \label{fig:systolic_array_block_diagram}
\end{figure}

Systolic arrays are the go-to architecture for the accelerated and efficient executions of general matrix multiplication (GeMM) operations, which are heavily employed by DNNs. Fig.~\ref{fig:systolic_array_block_diagram} shows the block diagram of the modeled systolic array architecture. It consists of a 2D grid structure of processing elements (PEs) responsible for performing scalar operations. Each PE communicates only with its adjacent neighbors in such a manner that data is passed only vertically down and horizontally right to the next PE. Data from the data memory can be loaded only into the first row and column of PEs and then flows through the systolic array. This is done by load units. The same applies to the last row and column of PEs to store results into the data memory by using store units.

\begin{figure}[tb]
    \centering
    \includegraphics[width=0.9\linewidth]{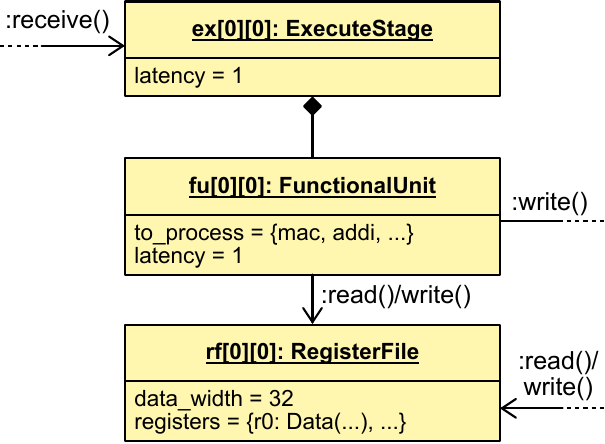}
    \caption{Architecture graph of the template for a PE of the parameterizable systolic array.}
    \label{fig:systolic_array_pe_template_ag}
\end{figure}

As already mentioned, the systolic array, more specifically the size of the PE grid, is parameterizable. This means that when creating the architecture, the parameters \texttt{rows} and \texttt{columns} are passed such that based on those parameters, the corresponding AG of the systolic array is instantiated. However, to accomplish this, statically instantiating all of the ACADL objects is not suitable anymore due to the parameter dependent amount of ACADL objects and edges. Therefore, we introduce templates which represent parts of the overall architecture that can be reused. This means instead of instantiating all ACADL objects and edges individually, we instantiate templates which create all the objects and edges belonging to the corresponding part of the architecture automatically. Listing~\ref{lst:systolic_array_pe_template_implementation} shows the implementation of the template representing a PE. Templates are implemented as Python classes in which all objects and edges belonging to the template are instantiated. Fig.~\ref{fig:systolic_array_pe_template_ag} shows the corresponding AG of the PE template. Templates allow us to model computer architectures using less code, facilitate the introduction of hierarchy, and increase reusability.

Every PE consists of a RegisterFile, an ExecuteStage and a FunctionalUnit, including the edges connecting these objects. One can also observe some edges without source or target in Listing~\ref{lst:systolic_array_pe_template_implementation}, we call these kinds of edges \textit{dangling edges}. Their purpose is to provide an interface for ACADL objects within a template to ACADL objects outside the template or inside other template instances. For example, each PE is connected with its adjacent neighbor and therefore edges between them are needed later on. For that reason, dangling edges keep the implementation more understandable and readable. When a dangling edge is not connected later on, no edge will be instantiated. To implement dangling edges, we use the class \texttt{ACADLDanglingEdge}.

\begin{lstlisting}[language=Python, caption=ACADL Python front-end implementation of the parameterizable systolic array PE template., label=lst:systolic_array_pe_template_implementation]
class ProcessingElement:
   def __init__(self, regs, row, col):
      # acadl objects
      self.ex = ExecuteStage(
         name=f"ex[{row}][{col}]", ...
      )
      self.fu = FunctionalUnit(
         name=f"fu[{row}][{col}]", ...
      )
      self.rf = RegisterFile(
         name=f"rf[{row}][{col}]", ...
      )

      # edges
      ACADLEdge(self.ex, self.fu, CONTAINS)
      ACADLEdge(self.rf, self.fu, READ_DATA)
      ACADLEdge(self.fu, self.rf, WRITE_DATA)

      # dangling edges
      self.ex_ingoing_forward = DanglingEdge(
         edge_type=FORWARD, target=self.ex
      )
      self.rf_ingoing_write = DanglingEdge(
         edge_type=WRITE_DATA, target=self.rf
      )
      self.rf_outgoing_read = DanglingEdge(
         edge_type=READ_DATA, source=self.rf
      )
      self.fu_outgoing_write = DanglingEdge(
         edge_type=WRITE_DATA, source=self.fu
      )
\end{lstlisting}

Listing~\ref{lst:systolic_array_pe_template_implementation} shows an example of how dangling edges are instantiated. One can either specify the source or the target of the edge, depending on the necessary edge later on, including the type of the edge. Later, dangling edges can be connected with each other by using the ACADL Python front-end function \texttt{connect\_dangling\_edge()}, whose output is an instance of \texttt{ACADLEdge}. It is also possible to pass a dangling edge and an ACADL object to this function, which might be helpful during the generation of the whole AG to maintain better readability or to avoid unnecessary instantiations of dangling edges outside a template. In either case, it is checked wether or not the resulting ACADL edge is valid with respect to the ACADL class diagram.

\begin{lstlisting}[language=Python, caption=ACADL Python front-end implementation of the parameterizable systolic array using templates and dangling edges., label=lst:systolic_array_implementation]
@generate
def generate_architecture(rows, columns)
    ...
    # instantiate array that holds all PEs
    pes = [
        [None]*columns for i in range(rows)
    ]
    ...
    # instantiate and connect PEs
    for row in range(rows):
        for col in range(columns):
            ...
            pes[row][col] 
                = ProcessingElement(
                    regs=4, row=row, col=col)
            ...
            connect_dangling_edge(
                pes[row-1][col]
                    .fu_outgoing_write, 
                pes[row][col]
                    .rf_ingoing_write
            )
            ...
        ...
    ...
\end{lstlisting}

With the help of templates and dangling edges, we can now generate the AG of the parameterizable systolic array in ACADL efficiently. Listing \ref{lst:systolic_array_implementation} shows how PE templates are instantiated and how \texttt{conntect\_dangling\_edge()} is used to connect instantiated templates. In addition to the template for the PE which has been presented in detail, we also implemented templates for the load units, store units and the fetch unit. The load and store units consist of an ExecuteStage and a MemoryAccessUnit which supports the load and store operations respectively. The fetch unit consists of the same objects and edges as already described in the OMA, including the instruction memory. Note that the DRAM object is not instantiated in any template. This will be done when generating the whole AG. It would be unnecessary to create a template for just one single ACADL object. In that case, it would also be unnecessary to instantiate a dangling edge, i.e. pass the DRAM object directly into the \texttt{connect\_dangling\_edge() function}.

Finally, the whole AG of the systolic array is composed by instantiating and connecting the templates and dangling edges. The parameters \texttt{rows} and \texttt{columns} can be set arbitrarily such that we can instantiate systolic array architecture of any desired shape.

\subsection{\goenna~ [gœna]}

\begin{figure}[tb]
    \centering
    \includegraphics[width=0.95\linewidth]{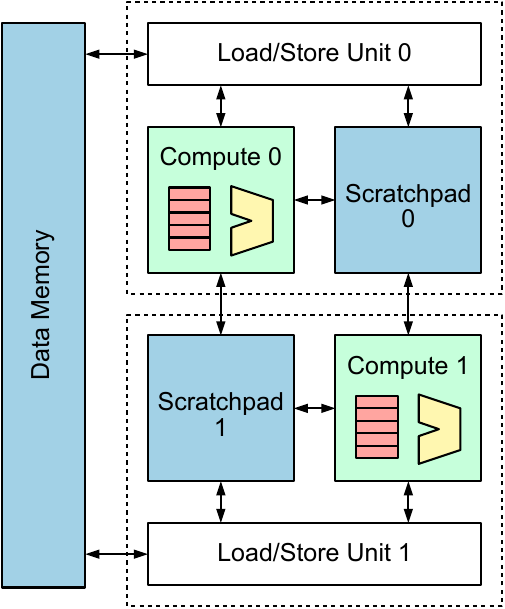}
    \caption{Block diagram of the \goenna~architecture, where the dashed lines enclose architecture templates.}
    \label{fig:goenna_block_diagram}
\end{figure}

In contrast to the OMA and the parameterizable systolic array which represent AI hardware accelerators modeled on the scalar operations level, the General Operationally Extendable Neural Network Accelerator (\goenna~[gœna]\footnote{While $\Gamma$ (Gamma) represents the generalizability, the second derivative of the position with respect to time is the acceleration: $\frac{d^2x}{dt^2} = \ddot{x}$}) is modeled on the fused tensor operations level. Fig.~\ref{fig:goenna_block_diagram} shows the block diagram of the \goenna~architecture. It is composed out of templates (enclosed in dashed lines) which contain a load/store unit, a compute unit, and a scratchpad. The compute units carry out fused tensor operations such as computing tiles of a general matrix multiplication (GeMM) with an optional activation function applied to the outputs or pooling while the matrix rows and columns are stored in vector registers. The scratchpad is an SRAM used to store partial results that can be shared with adjacent compute units. The load/store units move input and output data of operations between the DRAM data memory and the compute unit or the scratchpad. Instructions intended for different hardware components are issued in parallel and executed out-of-order, which facilitates an efficient and parallel computation of AI workloads.

\begin{lstlisting}[caption=\goenna~ACADL instruction for an 8$\times$8 matrix multiplication where the input matrices are loaded from a scratchpad into the the compute unit vector registers and the output matrix of the \texttt{gemm} instruction is stored in the scratchpad (the \texttt{1} parameter of the \texttt{gemm} instruction enables the ReLU activation function applied to the output matrix)., label=lst:goenna_instruction_example]
load  [0x3000]          => r[0].0
load  [0x3030]          => r[0].1
...
load  [0x3080]          => r[0].8
load  [0x4000]          => r[0].9
...
load  [0x4080]          => r[0].15
gemm  r[0].0, r[0].9, 1 => r[0].16 ; 1: ReLU
store r[0].16           => [0x5000]
...
store r[0].23           => [0x5080]
\end{lstlisting}

\begin{figure}[tb]
    \centering
    \includegraphics[width=\linewidth]{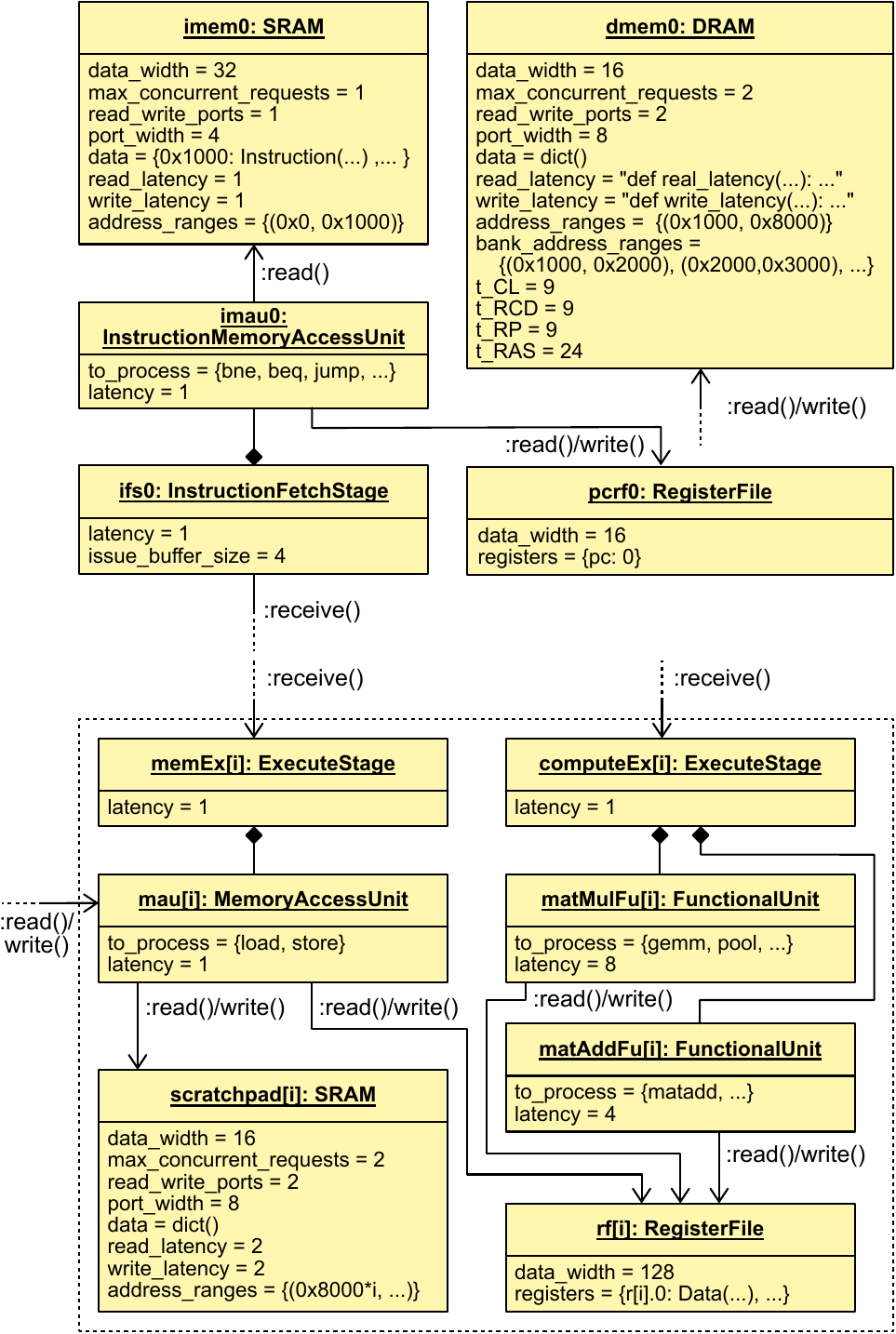}
    \caption{Architecture graph of the \goenna~architecture using a template for a compute and scratchpad pair.}
    \label{fig:goenna_object_diagram}
\end{figure}

Fig.~\ref{fig:goenna_object_diagram} presents the architecture graph of \goenna~using a template for the load/store, compute, and scratchpad complex. the \texttt{to\_process} attribute of the \texttt{matMulFu} and \texttt{matAddFu} contains the fused tensor operations that are carried out by those FunctionalUnits. E.g. the \texttt{gemm} operation calculates the matrix multiplication of two 8$\times$8 matrices with an optional activation such as rectified linear unit (ReLU) function applied to the output matrix using 16bit integer elements stored in 128bit wide vector registers. Listing~\ref{lst:goenna_instruction_example} shows an example for an 8$\times$8 matrix multiplication where the input matrices are loaded from the scratchpad into the compute unit vector registers and the result matrix from the \texttt{gemm} instruction is stored in the scratchpad.

Those three AI hardware accelerator modeling examples show the versatility of ACADL. Using templates and dangling edges in the Python front-end allows us to model computer architectures in an object-oriented fashion, while modeling at different abstraction levels enables fast prototyping and iterative refinement with explicit hardware parallelism.

\section{Operator Mapping}
\label{sec:operator_mapping}
To infer performance characteristics for an AI hardware accelerator modeled in ACADL, a DNN operation schedule is needed. As mentioned above, ACADL allows us to model computer architectures on different abstraction levels, from scalar operations up to fused tensor operations. We propose the use of Apache TVM \cite{tvm2018} and the Universal Modular Accelerator Interface (UMA) \cite{tvmuma2022} to map operations from a DNN onto the accelerator model. "Apache TVM is an open-source machine learning compiler framework for CPUs, GPUs, and machine learning accelerators. [\dots]" \cite{tvmwebsite2023} that facilitates code generation to offload typical machine learning workloads onto specialized hardware for accelerated execution. 

This section gives an overview of how a tiled general matrix multiplication (GeMM) is implemented on the OMA ACADL model and how we expose this tiled GeMM operator to TVM using the UMA interface.

\begin{figure}[ht]
    \centering
    \includegraphics[width=\linewidth]{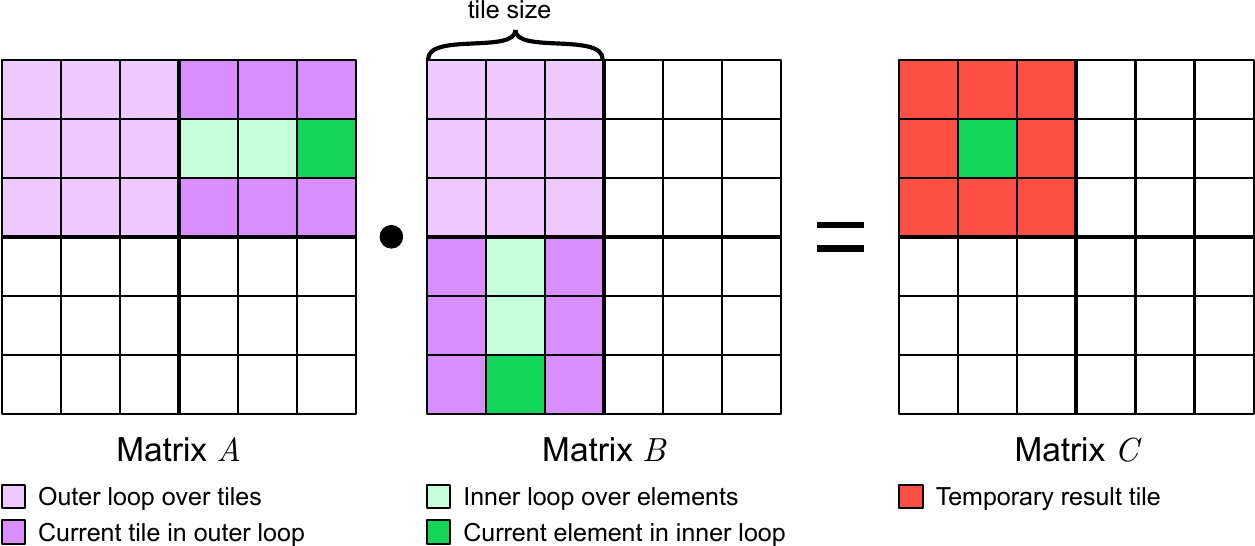}
    \caption{Illustration of tiled GeMM showing how single output matrix elements are computed. Purple tiles are inputs required to compute the red output tile, while the green row of input elements is processed to obtain the green output element. \cite{gemm2017}} 
    \label{fig:tiled_gemm}
\end{figure}

The OMA ACADL model is modeled on the scalar operations level. This means we need to implement the tiled GeMM operation with scalar operations represented by ACADL instructions. Listing \ref{lst:gemm_example} constitutes a naive example implementation of GeMM as a list of OMA ACADL instructions. Most notable among these instructions is the built-in multiply-and-accumulate (\texttt{mac}) instruction, which multiplies the contents of two input registers and adds them onto an accumulator register.
Regarding caching and locality, we want to offer various ways to compute the GeMM operation, e.g. via employing tiled matrix multiplication. This method utilizes a divide-and-conquer approach for matrix multiplication, which is based on partitioning the input matrices into submatrices called tiles, which are then multiplied and added:
\begin{align}
    \begin{bmatrix} A_{00} & A_{01} \\ A_{10} & A_{11} \end{bmatrix} \begin{bmatrix} B_{00} & B_{01}  \\ B_{10} & B_{11}\end{bmatrix} &= \begin{bmatrix}C_{00} & C_{01}  \\ C_{10} & C_{11}\end{bmatrix} \label{eqn:tiled_gemm}\\
    \text{where} \;\; A_{00} B_{00} + A_{01} B_{10}  &= C_{00} \label{eqn:tiled_gemm_start}\\
    A_{00} B_{01} + A_{01} B_{11} &= C_{01} \\
    A_{10} B_{00} + A_{11} B_{10} &= C_{10} \\
    A_{10} B_{01} + A_{11} B_{11} &= C_{11}. \label{eqn:tiled_gemm_end}
\end{align}
The above equation set details the tiled computation of the multiplication of matrices $A$ and $B$, divided into four tiles $A_{ij}$ and $B_{ij}$ as shown in equation~(\ref{eqn:tiled_gemm}). Equations~(\ref{eqn:tiled_gemm_start}) through (\ref{eqn:tiled_gemm_end}) detail which input tiles must be multiplied to obtain which output tile. Fig.~\ref{fig:tiled_gemm} visualizes the computation of a single output value of matrix $C$ using a tiled GeMM.

There are various execution orders in which a tiled GeMM can be computed, which has a significant impact on the execution time depending on the target computer architecture \cite{gemm2017}. For example, first computing $A_{00}B_{00}$ and $A_{00}B_{01}$ before moving on in a similar way, we only have to load the elements of $A_{00}$ into the target architecture's cache once for both sub-computations.

In the case of the OMA, we offer a mapping for a parameterizable implementation of the tiled GeMM. To be able to use this tiled GeMM mapping on the OMA accelerator, we propose the usage of the Universal Accelerator Interface (UMA) of TVM. With UMA, accelerator architectures can be easily integrated into the TVM DNN compilation flow by registering an interface function which implements a DNN operator such as GeMM. This GeMM interface function expects the input matrices along with their shapes, as well as the chosen tile size as arguments. To accelerate, e.g. a convolution operation, one needs to define the necessary input data transformations and computation schedules such that TVM can match and transform the convolution operations and insert the calls to the accelerator interface function into a C program.

\begin{lstlisting}[caption={OMA ACADL instruction list example showing an implementation of GeMM. Registers \texttt{r0} , \texttt{r1} and \texttt{r2} hold dimensions $m$, $n$ and $l$ of the input matrices $A$ ($m \times n$) and $B$ ($n \times l$). These indices are decremented until they reach zero, and then possibly reset in case of $n$ and $l$. The reset values for $n$ and $l$ are held in registers \texttt{r3} and \texttt{r4} respectively. Registers \texttt{r9} and \texttt{r10} hold the current input matrix element addresses, while \texttt{r11} holds the currently computed output matrix element's address. The initial input matrix addresses, intended as reset values for \texttt{r10} and \texttt{r11}, are held in registers \texttt{r12} and \texttt{r13}.}, label=lst:gemm_example]
mov   z0          => r8
load  [r9]        => r6
load  [r10]       => r7
mac   r6, r7      => r8
addi  r3, #-1     => r3
beqi  r3, z0, #16 => pc
add   r9, r14     => r9
add   r10, r14    => r10
jumpi #-28
beqi  r4, z0, #64 => pc
store r8          => [r11]
mov   r1          => r3
beqi  r4, z0, #20 => pc
mov   r12         => r9
add   r10, r14    => r10
add   r11, r14    => r11
jumpi #-64
addi  r5, #-1     => r5
mov   r2          => r4
beqi  r5, z0, #24 => pc
add   r9, r14     => r9
mov   r13         => r10
add   r11, r14    => r11
mov   r9          => r12
jumpi #-96
\end{lstlisting}

In the case of an AI hardware accelerator modeled in ACADL such as the OMA, the interface function for GeMM \texttt{oma\_tiled\_gemm(\dots)} may generate ACADL instructions  as presented in Listing~\ref{lst:gemm_example} according to the arguments passed, and then runs a functional and optional timing simulation to validate the DNN operator mapping and infer performance results.

\section{Timing Simulation Semantics}
Section \ref{sec:acadl} already gives a brief overview of the timing semantics of ACADL, this section provides a detailed description of the ACADL timing simulation semantics. These semantics have been implemented in \cite{acadl2022} using an Architectural Instruction Dependency Graph (AIDG) for fast performance estimation of DNN operators mapped onto the systolic array architecture introduced in section \ref{sec:moodeling_examples}, an Eyeriss v1 \cite{eyerissv12017} derived accelerator, and a Plasticine derived architecture \cite{plasticine2017} using a fixed point analysis of consecutive loop iterations. 

To infer performance characteristics, an AG and ACADL instructions for a modeled computer architecture are necessary. When initializing a timing simulation, each ACADL object in the AG which has the \texttt{latency} attribute is initialized with an integer variable \texttt{t:=0} representing a latency counter in clock cycles and a boolean variable \texttt{ready:=True} indicating the readiness of a hardware module. The overall simulation is initialized with an integer variable \texttt{T:=0} representing the simulation time. We visualize the behavior of selected ACADL classes using state diagrams \cite{statecharts1987}. Nodes in a state diagram represent different states, while edges signify a state transition. The edge label contains the condition under which a state transition occurs (before /) and the actions triggered by a state transition (after /). All presented state transitions in this paper can only occur at the end of a clock cycle when \texttt{T} is incremented. 

At \texttt{T==0} the InstructionFetchStage starts fetching instructions from the instruction memory through its contained InstructionMemoryAccessUnit. The number of instructions fetched is equal to the \texttt{port\_width} of the instruction memory. The InstructionFetchStage continuously fetches instructions and stalls when the issue buffer has no space left for \texttt{port\_with} instructions. After the instructions have been loaded from memory, they reside in the issue buffer and are forwarded to connected and ready PipelineStages. Multiple instructions can be forwarded out-of-order at the same time, which facilitates the simulation of parallel computer architectures. Fig.~\ref{fig:state_diagram_instruction_fetch_stage} shows the state diagram of the InstructionFetchStage.

\begin{figure}[tb]
    \centering
    \begin{tikzpicture}[scale=0.9, every node/.style={scale=0.9}]
        \node[state, initial] at (0,8) (fetch) {fetch};
        \node[state] at (0,0) (stall) {stall};
        \node at (2,9) [sloped, align=center] {$\bigvee_{\texttt{ps}}$\;\texttt{ps.ready()}\\/\texttt{ps.forward(),}\\\texttt{insts:=insts-1}};
        \node at (0,11.5) [sloped, align=center] {\texttt{issue\_buffer\_has\_space = }\\ \texttt{insts+imem.port\_width}\\ \texttt{<= issue\_buffer\_size}};
        \draw   (fetch) edge[bend left] node[sloped, align=center, below, rotate=180] {\texttt{imau.ready()} $\wedge$ \texttt{!issue\_buffer\_has\_space}\\/\texttt{insts:=insts+imem.port\_width}} (stall)
                (stall) edge[bend left] node[sloped, align=center, above] {\texttt{imau.ready()} $\wedge$ \texttt{issue\_buffer\_has\_space}\\/\texttt{insts:=insts+imem.port\_width}} (fetch)
                (stall) edge[loop below, double] node [sloped, align=center] {$\bigvee_{\texttt{ps}}$\;\texttt{ps.ready()}\\/\texttt{ps.forward(),insts:=insts-1}} (stall)
                (fetch) edge[loop above] node[sloped, align=center, above] {\texttt{imau.ready()} $\wedge$ \texttt{issue\_buffer\_has\_space}\\/\texttt{insts:=insts+imem.port\_width}} (fetch)
                (fetch) edge[loop right, double] node[sloped, align=center, above] {} (fetch)
        ;
    \end{tikzpicture}
    \caption{State diagram of the ACADL InstructionFetchStage class where \texttt{imau} is the contained InstructionMemoryAccessUnit object, \texttt{imem} is the instruction memory object, \texttt{ps} stands for any connected PipelineStage, and \texttt{insts} is the number of instructions residing in the issue buffer. The double arrow signifies that multiple instructions can be forwarded out-of-order to PipelineStages in the same clock cycle.}
    \label{fig:state_diagram_instruction_fetch_stage}
\end{figure}
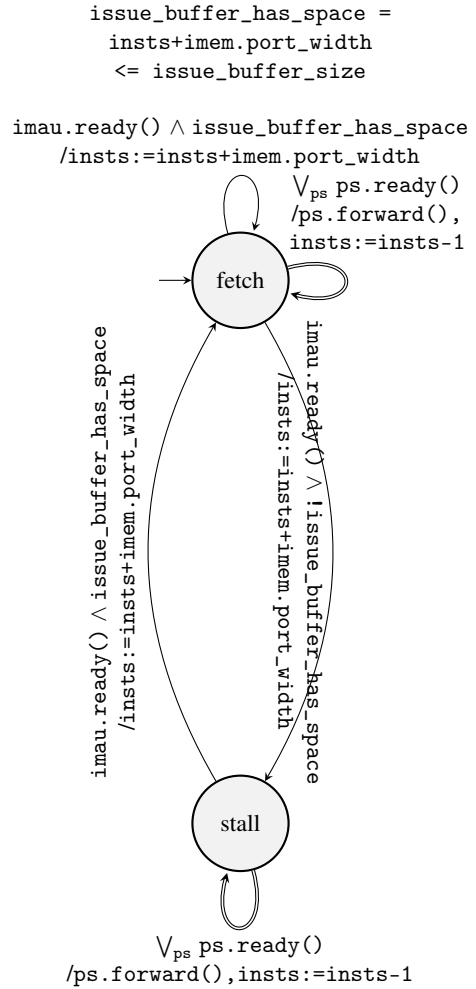

\begin{figure}[tb]
    \centering
    \begin{tikzpicture}[scale=0.9, every node/.style={scale=0.9}]
        \node[state, initial] at (0,6.5) (idle) {idle};
        \node[state] at (0,12.5) (buffer) {buffer};
        \node[state] at (0,0) (process) {process};
        \node[state] at (5.5,6.5) (stall) {stall};
        \draw   (idle) edge node[sloped, anchor=center, align=center] {\texttt{:receive()} $\wedge$ no \texttt{fu} found\\/\texttt{t:=latency,ready:=False}} (buffer)
                (buffer) edge[bend right] node[sloped, align=center, above, rotate=180] {\texttt{t==0} $\wedge$ \texttt{ps.ready()}\\\texttt{/:forward(ps),ready:=True}} (idle)
                (idle) edge node[sloped, anchor=center, align=center, rotate=180] {\texttt{:receive()} $\wedge$ \texttt{fu} found\\/\texttt{fu.process(),}\texttt{ready:=False}} (process)
                (process) edge[bend left] node[sloped, anchor=center, align=center,above] {\texttt{fu.ready()}\\/\texttt{ready:=True}} (idle)
                (buffer) edge node[sloped, anchor=center, above, align=center] {\texttt{t==0} $\wedge$ \texttt{!ps.ready()/}} (stall)
                (stall) edge node[align=center, below] {\texttt{ps.ready()}\\/\texttt{:forward(ps),}\\\texttt{ready:=True}} (idle)
                %(idle) edge[in=30,out=60,loop] node {} (idle)
                (stall) edge[loop above] node {} (stall)
                (buffer) edge[loop above] node {\texttt{t!=0}/\texttt{t:=t-1}} (buffer)
                (process) edge[loop below] node {\texttt{!fu.ready()}/} (process)
        ;
    \end{tikzpicture}
    \caption{State diagram of the ACADL ExecuteStage class, where \texttt{fu} represents any of the contained FunctionalUnits, and \texttt{ps} represents any of the connected PipelineStages.}
    \label{fig:state_diagram_execute_stage}
\end{figure}

When an ExecuteStage receives an instruction, it is checked if any of the contained FunctionalUnits can process it. If a FunctionalUnit is found which supports the instruction, the ExecuteStage sends the instructions to the FunctionalUnit and waits until the processing is finished. If no FunctionalUnit is found, the instruction is buffered for \texttt{latency} cycles and afterward forwarded to a ready and connected PipelineStage. While buffering an instruction or waiting on a FunctionalUnit processing, the ExecuteStage is not ready and cannot receive new instructions, which facilitates the simulation of structural hazards. The behavior of a PipelineStage is identical to the behavior of an ExecuteStage without any contained FunctionalUnits. Fig.~\ref{fig:state_diagram_execute_stage} presents the state diagram for the ExecuteStage class.

When a FunctionalUnit or MemoryAccessUnit receives an instruction, it becomes unready. Before processing an instruction in \texttt{latency} clock cycles, it is checked if all previous in-order instructions modifying the accessed registers and memory addresses are finished. This is facilitated by a global hash map which contains the last user for each register and memory address and ensures correct simulation of data dependencies. Fig~\ref{fig:state_diagram_memory_access_unit} visualizes the behavior of the MemoryAccessUnit and FunctionalUnit classes in a state diagram.

\begin{figure}[tb]
    \centering
    \begin{tikzpicture}[scale=0.9, every node/.style={scale=0.9}]
        \node[state, initial] at (0,6) (idle) {idle};
        \node[state] at (0,0) (process) {process};
        \draw   (idle) edge[bend left] node[sloped, align=center, below, rotate=180] {\texttt{:process()}\\/\texttt{t:=latency,ready:=False}} (process)
                (process) edge[bend left] node[sloped, align=center, above] {\texttt{t==0}\\/\texttt{ready:=True}} (idle)
                (process) edge[loop left] node [sloped, align=center, rotate=-90] {all previous\\data users have finished\\/\texttt{t:=t-1},\\update last data users} (process)
                %(process) edge[loop right] node {not all data avail./} (process)
                %(idle) edge[loop right] node {/} (idle)
        ;
    \end{tikzpicture}
    \caption{State diagram of the ACADL FunctionalUnit and MemoryAccessUnit classes.}
    \label{fig:state_diagram_memory_access_unit}
\end{figure}

\begin{figure}[tb]
    \centering
    \begin{tikzpicture}[scale=0.9, every node/.style={scale=0.9}]
        \node[state, initial] at (2,6) (idle) {idle};
        \node[state] at (0,0) (read) {read};
        \node[state] at (4,0) (write) {write};
        \draw   (idle) edge[bend right] node[sloped, above, align=center] {\texttt{:read()}\\/\texttt{t:=read\_latency,ready:=False}} (read)
                (read) edge node[sloped, align=center, below] {\texttt{t==0}/\texttt{ready:=True}} (idle)
                (read) edge[loop below] node {\texttt{t!=0}/\texttt{t:=t-1}} (read)
                (idle) edge[bend left] node[sloped, above, align=center] {\texttt{:write()}\\/\texttt{t:=write\_latency,ready:=False}} (write)
                (write) edge node[sloped, below] {\texttt{t==0}/\texttt{ready:=True}} (idle)
                (write) edge[loop below] node {\texttt{t!=0}/\texttt{t:=t-1}} (write)
        ;
    \end{tikzpicture}
    \caption{State diagram of the ACADL MemoryInterface request slot.}
    \label{fig:state_diagram_memory_interface}
\end{figure}
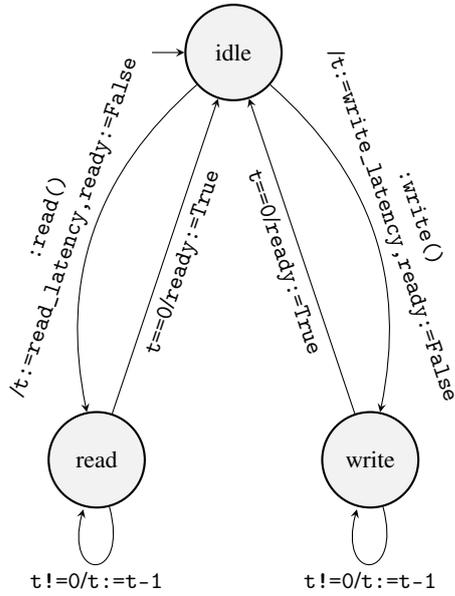

Classes inheriting from the virtual DataStorage class support up to \texttt{max\_concurrent\_requests} requests at the same time. Each request slot has its own \texttt{t} and \texttt{ready} variables. When \texttt{read()} or \texttt{write()} is called, it is checked if any request slot is ready. If no request slot is ready, the access is buffered  in a FIFO queue and is assigned to the next request slot that becomes ready. For classes inheriting from \texttt{MemoryInterface} Fig.~\ref{fig:state_diagram_memory_interface} shows the state diagram. When data is read from or written to an address, the latency counter \texttt{t} is initialized with the respective latency value, which is either static or can be provided by a memory simulator such as DRAMsim3 \cite{dramsim32020}.

\begin{figure}[!ht]
    \centering
    \begin{tikzpicture}[scale=0.9, every node/.style={scale=0.9}]
        \node[state, initial] at (3,12) (idle) {idle};
        \node[state] at (0,6) (read miss) {\shortstack{read\\miss}};
        \node[state] at (0,0) (read hit) {\shortstack{read\\hit}};
        \node[state] at (6,6) (write miss) {\shortstack{write\\miss}};
        \node[state] at (6,0) (write hit) {\shortstack{write\\hit}};
        \draw   (idle) edge[bend right] node[sloped, align=center, above] {\texttt{:read()} $\wedge$ address not in cache\\/\texttt{t:=miss\_latency,ready:=False}} (read miss)
                (read miss) edge[in=30,out=60,loop] node[sloped, align=center, above, rotate=45] {\texttt{t!=0}\\/\texttt{t:=t-1}} (read miss)
                (read miss) edge node[sloped, align=center, above, rotate=180] {\texttt{t==0}\\/\texttt{t:=hit\_latency,}\\\texttt{:update\_cache()}} (read hit)
                (read hit) edge[loop right] node[sloped, align=center, rotate=90] {\texttt{t!=0}\\/\texttt{t:=t-1}} (read hit)
                (read hit) edge node[sloped, above] {\texttt{t==0}/\texttt{ready:=True}\quad\quad\quad\quad\quad\quad} (idle)
                %(idle) edge[bend left] node {} (read hit)
                (idle) edge[bend left] node[sloped, align=center, above] {\texttt{:write()} $\wedge$ address no in cache\\/\texttt{t:=miss\_latency,ready:=False}} (write miss)
                (write miss) edge node[sloped, align=center, below, rotate=180] {\texttt{t==0}\\/\texttt{t:=hit\_latency,}\\\texttt{:update\_cache()}} (write hit)
                (write miss) edge[in=135,out=105,loop] node[sloped, align=center, above, rotate=-30] {\texttt{t!=0}\\/\texttt{t:=t-1}} (write miss)
                (write hit) edge[loop left] node [sloped, align=center, rotate=-90] {\texttt{t!=0}\\/\texttt{t:=t-1}} (write hit)
                (write hit) edge node[sloped, align=center, below] {\texttt{t==0}/\texttt{ready:=True}} (idle)
                (idle) edge[bend left=12] node[sloped, align=center, below, right=-5] {\vphantom{H}\\\vphantom{H}\\\texttt{:read()} $\wedge$ address in cache\\/\texttt{t:=hit\_latency,}\texttt{ready:=False}} (read hit)
                (idle) edge[bend right=12] node[sloped, align=center, below, left=-5] {\vphantom{H}\\\vphantom{H}\\\texttt{:write()} $\wedge$ address in cache\\/\texttt{t:=hit\_latency,}\texttt{ready:=False}} (write hit)
        ;
    \end{tikzpicture}
    \caption{State diagram of the ACADL CacheInterface request slot.}
    \label{fig:state_diagram_cache_interface}
\end{figure}
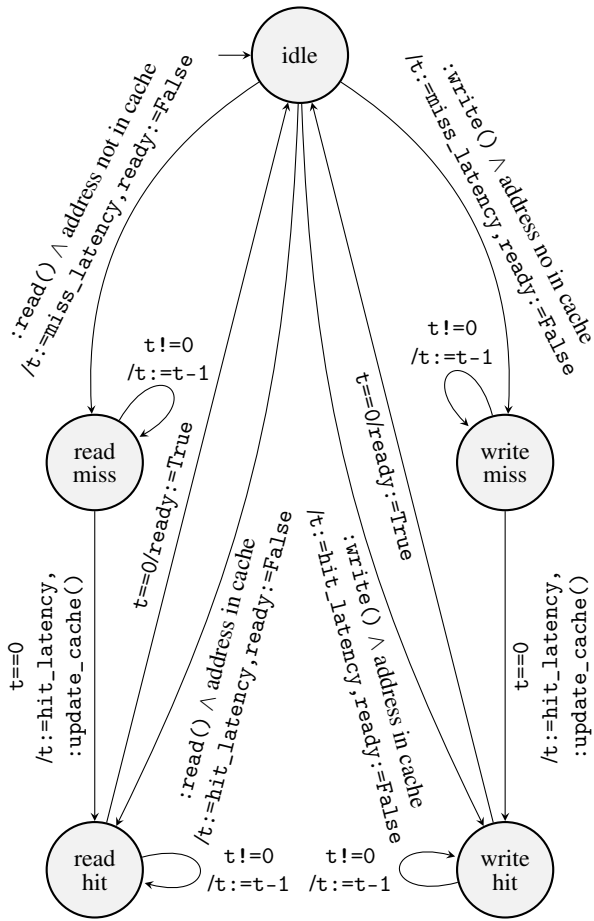

The timing semantics for classes inheriting from CacheInterface is very similar to classes inheriting from MemoryInterface. Each request slot has its own set of variables and requests are processed in a FIFO fashion. In addition to classes inheriting from MemoryInterface it is checked if an accessed address resides in the cache, which is facilitated by using a cache simulator such as pycachesim \cite{pycachesim2017}. When a miss occurs, \texttt{t} is set to the \texttt{miss\_latency}. After \texttt{miss\_latency} clock cycles have passed, the cache simulator is updated and \texttt{t} is set to the \texttt{hit\_latency} and the request slot becomes ready again after \texttt{miss\_latency} clock cycles have passed. Fig.~\ref{fig:state_diagram_cache_interface} details the behavior of a CacheInterface request slot.

The above ACADL timing semantics have been implemented in \cite{acadl2022} using an Architectural Instruction Dependency Graph (AIDG) which enables fast performance estimation of DNN operators.

\section{Conclusion}
This paper presented how ACADL can be used to model AI hardware accelerators and outlines how DNN operators can be mapped to these modeled computer architectures using TVM, along with an introduction to the timing simulation semantics of ACADL. This allows researchers and engineers to quickly evaluate AI hardware accelerator designs on different abstraction levels, from scalar operations up to fused tensor operations. Additionally, the timing simulation can be used in the optimization loop of hardware-aware NAS and DNN/HW Co-Design to find close to optimal design points.

\section{Acknowledgement}
This work has been funded by the German Federal Ministry of Education and Research (BMBF) under grant number 16ES0876 (GENIAL!).

\bibliographystyle{IEEEtran}
\bibliography{references}

\end{document}